\documentclass[aip,graphicx,twocolumn,10pt, apl]{revtex4-1}
\usepackage{graphicx}
\usepackage{epstopdf}
\usepackage{afterpage}
\usepackage{times}

\begin{document}

%\preprint{}

\title{Fabrication and Characterization of Topological Insulator Bi$_2$Se$_3$ Nanocrystals} %Title of paper



\author{ S.Y.F. Zhao$^1$, C. Beekman$^1$,  L.J. Sandilands$^1$, J.E.J. Bashucky$^1$, D. Kwok$^2$, N. Lee$^2$,  A.D. LaForge$^3$ ,S.W. Cheong$^2$ and  K.S. Burch} \email{kburch@physics.utoronto.ca}
%\thanks{}
\affiliation{$^1$Department
of Physics \& Institute of Optical Sciences, University of
Toronto, 60 St. George Street, Toronto, ON M5S 1A7\\
$^2$Rutgers Center for Emergent Materials and Department of Physics and Astronomy, Rutgers University, 136 Frelinghuysen Road, Piscataway, NJ 08854, USA.\\
$^3$Department of Physics, University of California, Santa Cruz, 1156 High Street, Santa Cruz, California 95064, USA.}


\date{\today}

\begin{abstract}
\noindent In the recently discovered class of materials known as topological insulators, the presence of strong spin-orbit coupling causes certain topological invariants in the bulk to differ from
their values in vacuum. The sudden change of invariants at
the interface results in metallic, time reversal invariant
surface states whose properties are
useful for applications in spintronics and quantum computation. However, a key challenge is to fabricate these materials on the nanoscale appropriate for devices and probing the surface. To this end we have produced 2 nm thick nanocrystals of the topological insulator Bi$_2$Se$_3$ via
mechanical exfoliation. For crystals thinner than 10 nm we observe the emergence of an additional mode
in the Raman spectrum. The emergent mode intensity together with
the other results presented here
provide a recipe for production and thickness characterization of Bi$_2$Se$_3$ nanocrystals. \\
%The topological invariants in the bulk of insulators can differ from the topological invariants in vacuum but can only be altered in the presence of time reversal symmetry breaking. Due to the presence of strong spin-orbit coupling There has been growing interest in insulators have such a gapless time-reversal invariant surface states (i.e. topological insulators).  The properties of these surface states can be useful for many applications in spintronics and quantum computation. However a key challenge is preparing these materials on the nano-scale appropriate for devices and primarily probing the surface. To this end we have produced single unit cell thick nanocrystals of the topological insulator Bi$_2$Se$_3$ via mechanical exfoliation. The nanocrystals are characterized by optical, Raman and atomic force microscopy. Surprisingly, for crystals thinner than 6 nm we observe the emergence of a new mode in the Raman spectrum. The emergent mode intensity together with the other results presented here provide a recipe for production and thickness determination of Bi$_2$Se$_3$ nanocrystals
\end{abstract}

\maketitle
\begin{figure}[h]
  \includegraphics[angle=0, width=5.7cm, height=8.3cm]{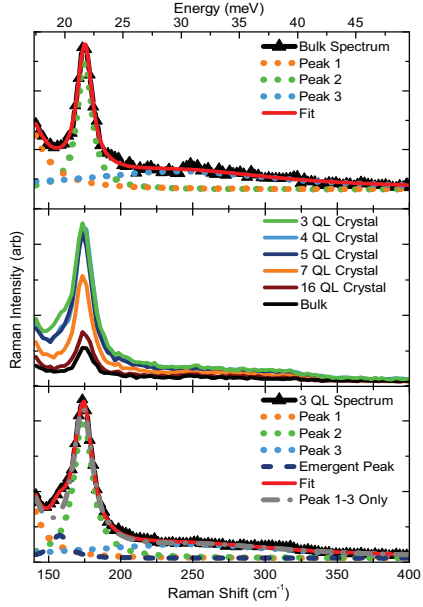}
  \caption{\footnotesize (Color online) Raman spectra for Bi$_2$Se$_3$ (a) Spectrum for the bulk crystal (line + symbols) with the corresponding fit (solid line) consisting of three Lorentzian oscillators (dotted lines). (b) Spectra for crystals of varying thicknesses, 3 $\rightarrow$16 QL and bulk. (c) Spectrum for 3QL nanocrystal. The data (line + symbols) is fit with four Lorentzian oscillators (solid line). The fit resulting from three oscillators is shown for comparison (dashed dotted line).}\label{fig3} 
  \end{figure}
\begin{figure}[h]
 \includegraphics[angle=0, width=8.6cm, height=3.8cm]{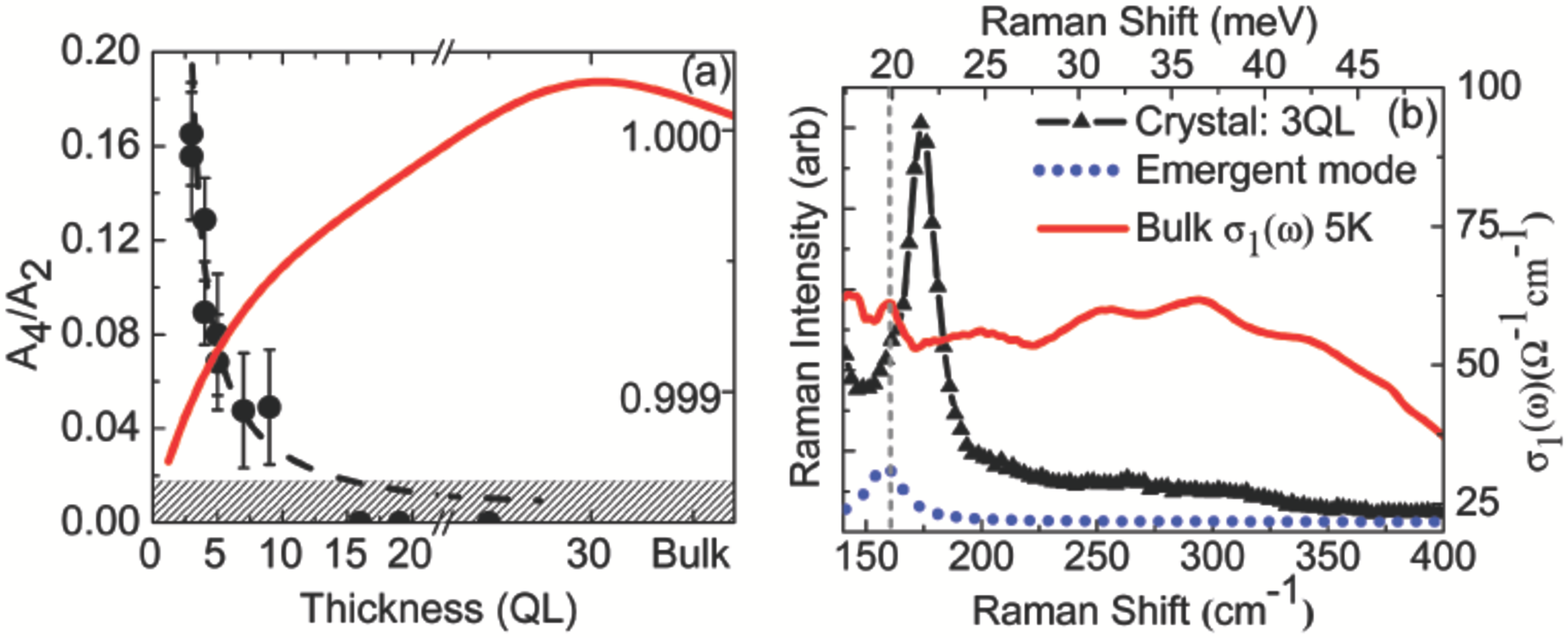}
 \caption{\footnotesize (Color online) (a) Ratio of emergent peak intensity to the main peak intensity (determined from fits) as function of crystal thickness. The dashed line is a guide to the eye and plotted to show the $\frac{\alpha}{(QL-2)}$ trend of the ratio. The gray area for A$_4$/A$_2 <$0.02 indicates the detection limit of our system. The ratio A$_4$/A$_2$ as function of crystal thickness due to FP effects (solid line, right axis).  (b) Left axis: Raman spectrum for a 3QL thick crystal (line + symbols) and the lorentzian corresponding to the emergent mode (dotted line). Right axis: optical conductivity data (ssolid line) of a bulk Bi$_2$Se$_3$ single crystal taken at T = 5 K.
 }\label{fig4}
\end{figure}
\begin{figure}[h]
 \includegraphics[angle=0, width=5.5cm, height=4.5cm]{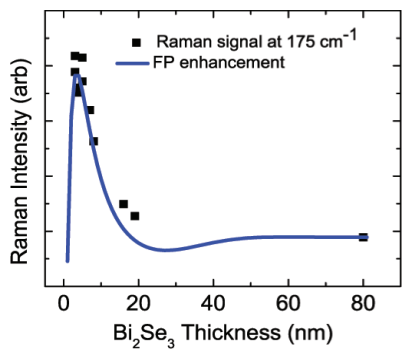}
  \caption{\footnotesize (Color online) Raman intensities as function of crystal thickness at 175 cm$^{-1}$ (squares) and  calculated Raman intensities due to FP interference effects (solid line: optical constants modified from bulk values). }\label{Figoninterference}
\end{figure}

%\end{abstract}
%\pacs{}% insert suggested PACS numbers in braces on next line
 %\maketitle must follow title, authors, abstract and \pacs
% References should be done using the \cite, Fig. \ref, and \label commands
%\section{}
%\label{}
%\subsection{}
%\subsubsection{}
Topological metallic surface states are predicted to have numerous properties that are useful for spintronics and quantum computation.\cite{moore, 2009NatPh...5..438Z}
A challenging aspect of this research has been to isolate surface state contributions to the measured properties of topological insulators \cite{Peng:2010p2379,Eto:2010p2376,PhysRevB.81.205407,Zhang:2009p2371,PhysRevB.81.041405}. Studying nanometer thick crystals allows one to tune the chemical potential with electric field\cite{2010ApPhL..96o3103S,
2010Nanolett_BiTe,b820226e,PhysRevB.81.165432,Checkelsky:2010p2307}, as well as observe modifications of the excitation spectrum produced by interactions of top and bottom surfaces\cite{PhysRevB.80.205401,PhysRevB.81.041307,PhysRevB.81.115407,Bi2Se3thinGap,PhysRevB.81.165432}. 
One option for producing thin crystals is mechanical exfoliation. This method, together with Raman spectroscopy, has proven to be
extremely fruitful in the study of
graphene\cite{Gupta:2006p2359,PhysRevLett.103.116804} and other nanocrystals \cite{Malard:2009p2363,luke}. Indeed, Raman provides direct access to the phonon spectrum and can be used to map the
thickness\cite{Gupta:2006p2359,Yoon:2009p2241}
or doping level\cite{Basko:2009p2367}
over a large area.\\\indent
Similar studies of the topological insulators Bi$_{2}$Se$_{3}$ and
Bi$_{2}$Te$_{3}$ have been attempted. To date, these
experiments have been limited to exfoliated Bi$_{2}$Se$_{3}$
crystals $>$10 nm thick\cite{Steinberg:2010p2380,Checkelsky:2010p2307, morpurgo} or
Bi$_{2}$Te$_{3}$ where the bulk gap is
small\cite{2010ApPhL..96o3103S, 2010Nanolett_BiTe}. A limiting
factor in these experiments was the strong optical absorption of
these compounds, making the identification of thin crystals
difficult on Si/SiO$_2$ substrates. With this in
mind, we have mechanically exfoliated Bi$_{2}$Se$_{3}$ crystals on
Mica, enabling us to optically identify crystals only 2 nm thick.
By systematically studying these crystals with Raman spectroscopy and
optical and atomic force microscopies (AFM) we have devised a method for
characterizing the thickness of Bi$_{2}$Se$_{3}$ nanocrystals using
non-invasive, all-optical methods. Specifically, an
additional mode appears in the Raman spectra for ultrathin ($<$  10 nm)
crystals. The observed thickness dependence of the emergent mode intensity
can be used for thickness verification of nanocrystals via Raman
measurements.\\\indent    
Bi$_2$Se$_3$ forms a rhombohedral lattice in which the unit cell is composed of
three five-layer stacks known as quintuple layers (QL). A unit
cell measures 2.87 nm along the $c$ axis, and 1 QL is $\approx$ 0.96 nm
thick.\cite{Richter:1977p2372} Atoms are arranged into
planar hexagonal sheets with the sequence
-[Se$^{(2)}$-Bi-Se$^{(1)}$-Bi-Se$^{(2)}$]- \cite{sup1}.
The superscripts indicate the structural in-equivalence of the
Se ions, with the Se$^{(1)}$ atom at a center of inversion within
a QL. Therefore one expects phonon modes to be exclusively
infrared (IR) or Raman active. The weak van der Waals
bonds between neighboring Se$^{(2)}$ planes enables mechanical exfoliation. Ultrathin Bi$_2$Se$_3$ nanocrystals could be identified after
exfoliation by optical microscopy in transmission mode. The crystal thickness was subsequently determined using AFM\cite{sup1}.\\\indent
Raman spectroscopy ($\lambda$ = 532 nm, spot size $\sim$ 1 $\mu$m) confirmed that these
nanocrystals are indeed Bi$_{2}$Se$_{3}$ and was used to study the
evolution of the crystal lattice structure with thickness. 
A raman spectrum ($I(\omega)$) for a bulk
crystal is shown in Fig. \ref{fig3}a, where we observe a strong
phonon mode at 175 cm$^{-1}$ and the onset of a mode below 150
cm$^{-1}$. Previous results\cite{Richter:1977p2372} suggest the higher energy mode
corresponds to a Raman active A$_{1g}$ mode and the latter to an
E$_g$ mode, in accord with group theory predictions for
phonons at the Brillouin zone center (\emph{q} = 0) probed by optical
experiments. Indeed, the 5 atoms in the unit cell should lead to
12 optical modes, each of which is either exclusively Raman or infrared active\cite{Richter:1977p2372}. We also observe a broad shoulder between
200-350 cm$^{-1}$, similar to a feature observed in a recent IR study of bulk
Bi$_2$Se$_3$ (see Fig.\ref{fig4}b).\cite{2010PhRvB..81l5120L} The presence of
this feature in both Raman as well as IR and its broad lineshape suggest that it is due to two-phonon excitations.
\\\indent
To quantitatively analyze the evolution of the spectra, we 
fit the measured Raman data with multiple Lorentzian oscillators in the form:
$I(\omega) = I_0 + \sum_i
(\frac{A_i\Gamma_{i}}{(4(\omega-E_{i})^2 +
\Gamma_{i}^2)})$ where $i$ ranges from 1 to 3 or 4,
depending on thickness, $I_0$ accounts for the background,
$E_{i}$ is the center, $\Gamma_{i}$ is the width, and $A_{i}$ is the area of peak $i$. The resulting fit
for the bulk spectrum with three oscillators is shown in Fig.
\ref{fig3}a, where the center frequency of the E$_g$ mode
(Peak 1) was fixed at 131.5 cm$^{-1}$ based on previous
studies\cite{Richter:1977p2372}. Turning to Fig. \ref{fig3}b we
examine the evolution of the Raman spectra with varying crystal
thicknesses, where an enhancement of the overall signal is observed with decreasing number of QL. As is
discussed below this enhancement results from multiple
reflections in the Bi$_2$Se$_3$ crystal. Perhaps more surprisingly, an additional mode
emerges at 158 cm$^{-1}$ as the crystal thickness is reduced (see Fig. \ref{fig3}c). Indeed, the use of
only three oscillators, which worked well for crystals thicker
than 10 QL (Fig. \ref{fig3}a), results in a large difference
around 158 cm$^{-1}$ between the fit (dashed dotted line) and the spectrum.
By simply adding another mode ($i=4$), the
fit (solid line) agrees very well with the data. 
The emergence of the mode at 158 cm$^{-1}$ for ultrathin crystals suggests it can be used to verify sample thickness.\\\indent We explore this possibility in Fig. \ref{fig4}, where we plot the ratio of the emergent mode intensity with the main peak intensity ($\frac{A_{4}}{A_{2}}$). For thicknesses below 10 QL this ratio increases and agrees well with $\frac{\alpha}{(QL-2)}$ behavior (dashed line), which corresponds to the relative weight of surface and bulk modes. Indeed,  the intensity of a particular mode is proportional to the volume over which the light can emanate. For a surface mode, this volume is independent of thickness. In contrast, for a bulk mode the volume probed is proportional to the crystal thickness (for thicknesses less than the penetration depth). %The amplitude of the emergent mode may provide a means for thickness verification of Bi$_2$Se$_3$ nanocrystals. 
Unfortunately, due to the detection limit of our instrument, we are not able to convincingly detect the mode for A$_4$/A$_2$ $<$ 0.02. %In Fig. \ref{fig4}b we show the Raman spectrum for the 3QL crystal, the corresponding lorentzian describing the emergent mode and optical conductivity data from a Bi$_2$Se$_3$ single crystal taken at T = 5 K\cite{2010PhRvB..81l5120L}. The optical conductivity shows an IR-phonon mode at the same energy as our emergent mode, which has become Raman active.
    \\\indent
While the mode appears to have its origin in the surface, it could also be due to Fabry-Perot (FP) interference in the Mica substrate. To rule this out, we repeatedly performed Raman spectroscopy on a single 3 QL nanocrystal and subsequently cleaved the back surface of the Mica to reduce its thickness. The resultant spectra all overlapped (not shown), implying that FP interference in the Mica substrate can be neglected. Nonetheless, for a fixed substrate thickness, the interference effects due to multiple reflections in the Bi$_2$Se$_3$ nanocrystals will change as the crystals are thinned. To check the FP effects on the Raman spectra we have performed a calculation similar to the one shown to work well in graphene and Bi$_2$Sr$_2$CaCu$_2$O$_8$ nanocrystals.\cite{Yoon:2009p2241,luke} In Fig. \ref{Figoninterference}, we plot the measured and the calculated intensity (FP model) as a function of crystal thickness for the main mode (175 cm$^{-1}$). Changes in the ratio ($\frac{A_{4}}{A_{2}}$) due to FP effects are plotted in Fig. \ref{fig4}a and only reveals a very small dependency on crystal thickness, which is opposite to what we observe. Therefore, the emergent mode is intrinsic to Bi$_2$Se$_3$ nanocrystals  and not caused by FP effects. However, Fig.\ref{Figoninterference}  shows that the change in the overall Raman signal with crystal thickness is explained by FP interference effects only if the optical constants are modified from the bulk values. This modification is not unreasonable given recent photoemission experiments\cite{Bi2Se3thinGap,PhysRevB.81.041307}. Interestingly, these data also show the utility of Raman measurements, as the overall intensity can be used to determine the thickness of the Bi$_2$Se$_3$ nanocrystals.
   \\\indent
While the FP interference described above can account for the overall intensity-thickness trends in Fig. \ref{fig3}, the origin of the 158 cm$^{-1}$ mode remains unclear. Interestingly, an additional mode also appeared in Raman spectra of nanocrystals of the isostructural topological insulator Bi$_2$Te$_3$.\cite{2010ApPhL..96o3103S} This mode matched perfectly the frequency of an infrared-active mode and so it was attributed to the breaking of inversion symmetry. Shahil \emph{et al} suggested that mechanical exfoliation resulted in breaks within a QL as well as between them. A  similar explanation may be appropriate for Bi$_{2}$Se$_{3}$: a recent IR reflectance study of the bulk revealed a mode with the same frequency and width\cite{2010PhRvB..81l5120L} (see Fig.\ref{fig4}b). However, in Bi$_2$Te$_3$ this mode appeared in crystals thinner than 84 nm, whereas it only appears in Bi$_2$Se$_3$ nanocrystals thinner than 10 nm. We believe the mode may be due to the built-in electric fields at the surface. Specifically, the band bending inherent to materials with surface states will generate asn electric field that will break the inversion symmetry.     %Furthermore, the self-consistency of the Raman results implies cleavage does not occur within a QL. It is interesting to note that the emergent mode appears for the same thickness of Bi$_2$Se$_3$ where coupling of the two surfaces results in a modification of the electronic structure.\cite{Bi2Se3thinGap,PhysRevB.81.041307}
\\\indent We have shown that Raman spectroscopy is an effective
nanometrology tool for identifying nanocrystals of the topological
insulator Bi$_2$Se$_3$. This is accomplished by monitoring i) the overall intensity of the Raman signal and/or ii) the strength of the emergent mode at 158 cm$^{-1}$. The overall thickness dependence of
the intensity can be accounted for by proper modeling of the
effect of interference on the Raman spectra. 
The origin of the emergent mode remains unclear, although the presented optical conductivity data suggests that inversion symmetry breaking leads to an IR mode becoming Raman active. However, it is interesting to note that the emergent mode appears for the same thickness regime of Bi$_2$Se$_3$ for which a gap has been theorized to open due to coupling of the two surfaces. \cite{PhysRevB.81.041307,Bi2Se3thinGap} In addition, the polar surface of Mica
may lead to band bending, which breaks inversion symmetry.
Therefore performing Raman spectroscopy on a suspended crystal
would provide conclusive determination of the influence of the
Mica substrate on the Bi$_2$Se$_3$ nanocrystals. Nonetheless, we
have provided a path for fabricating and identifying Bi$_2$Se$_3$
nanocrystals through the combination of mechanical exfoliation on
transparent substrates and the use of Raman spectroscopy. This work paves
the way for future devices  and studies of the surface states of
topological insulators.\\\indent
We are grateful for numerous discussions with Y.B. Kim and H.Y. Kee and we thank Y. J. Choi for the transport measurement. Work at the University of Toronto was supported by NSERC, CFI, and ORF; work at Rutgers University was supported by the NSF under grant NSF-DMR-0804109.

\end{document}